\documentclass[10pt]{revtex4}
\usepackage{amssymb}
\usepackage{latexsym}
\usepackage{epsfig}

\usepackage{amsmath}
\begin{document}
\title{Revisiting Agegraphic Dark Energy in Brans-Dicke Cosmology}
\author{M. Abdollahi Zadeh$^{1}$\footnote{m.abdollahizadeh@shirazu.ac.ir} and A.
Sheykhi$^{1,2}$\footnote{asheykhi@shirazu.ac.ir}}

\address{$^1$ Physics Department and Biruni Observatory, College of
Sciences, Shiraz University, Shiraz 71454, Iran\\
$^2$ Research Institute for Astronomy and Astrophysics of Maragha
(RIAAM), P.O. Box 55134-441, Maragha, Iran}

\begin{abstract}
We explore a spatially homogeneous and isotropic
Friedmann-Robertson-Walker (FRW) universe which is filled with
agegraphic dark energy (ADE) with mutual interaction with
pressureless dark matter in the background of Brans-Dicke (BD)
theory. We consider both original and new type of agegraphic dark
energy (NADE) and further assume the sign of the interaction term
can change during the history of the Universe. We obtain the
equation of state parameter, the deceleration parameter and the
evolutionary equation for the sign-changeable interacting ADE and
NADE in BD theory. We find that, in both  models, the equation of
state parameter, $w_D$, cannot cross the phantom line, although
they can predict the Universe evolution from the early
deceleration phase to the late time acceleration, compatible with
observations. We also investigate the sound stability of these
models and find out that both models cannot show a signal of
stability for different model parameters.
\end{abstract}

\maketitle
\section{Introduction\label{Int}}
Cosmological probes such as type Ia Supernova
\cite{Riess,Riess1,Riess2}, Weak Lensing \cite{Leauthand2010},
Cosmic Microwave Background (CMB) anisotropies
\cite{HAN2000,HAN20001}, Large-Scale Structure (LSS)
\cite{COL20011,COL20012,COL20013}, Plank data \cite{Ade2014} and
Baryon Acoustic Oscillations (BAO) \cite{Tegmark2004}, have given
us cross-checked data to determine cosmological parameters with
high precision. Combining the analysis of cosmological
observations we realize that our observable Universe is nearly
flat, homogeneous and isotropic at large scale and is currently
experiencing a phase of accelerated expansion. Besides, a phase
transition from deceleration to the acceleration was occurred in
the redshift around $0.45\leq z\leq 0.9$ \cite{Cai2010,Bamba}.

A great variety of scenarios have been proposed to explain this
acceleration such as some attempts to investigate the nature of
dark energy according to some principles of quantum gravity,
although a complete theory  of quantum gravity has not established
yet. The ADE model is such an example, which is based on the
uncertainty relation of quantum mechanics together with the
gravitational effects in general relativity. In this model it is
assumed that the observed dark energy comes from the quantum
fluctuations of the space time \cite{Cai20071,Cai20072}. In Refs.
\cite{Karolyhazy,Karolyhazy1,Karolyhazy2}, Karolyhazy and his
collaborators showed that the distance $t$ in Minkowski space time
cannot be known to a better accuracy than  $\delta{t}=\beta
t_{p}^{2/3}t^{1/3}$ where $\beta$ is a dimensionless constant of
order unity and $t_{p}$ denotes reduced Plank time. Based on
karolyhazy relation together with the time-energy uncertainty
relation, Maziashvili \cite{Maziashvili1,Maziashvili2} and
Sasakura \cite{Sasakura} have independently obtained the energy
density of spacetime fluctuations as
\begin{equation}\label{rho0}
\rho_{D} \sim \frac{1}{t_{p}^2 t^2} \sim \frac{m^2_p}{t^2},
\end{equation}
where $m_p$ and $t$ are the reduced Plank mass and proper time
scale, respectively. In the following, Cai \cite{Cai20071}
proposed the energy density of the original ADE has the form
\begin{equation}\label{rhoD}
\rho_{D}=\frac{3n^2m^2_p}{T^2}=\frac{3n^2}{8 \pi G
T^2},
\end{equation}
where $T$ is the age of the universe and the numerical factor
$3n^2$ is introduced to parameterize some uncertainties, such as
the species of quantum fields in the universe, the effects of
curved spacetime, etc. Since the original ADE model suffers from
the difficulty to describe the matter-dominated epoch, for
avoiding these  internal in consistencies, the NADE model was
proposed by Wei and Cai \cite{Cai20072}, by replacing the cosmic
age $T$ with  the cosmic conformal age $\eta$ for the time scale.
The ADE models have been studied extensively and constrained by
various astronomical observations \cite{Wei}. On the other side,
it is interesting to analyze both ADE and NADE models in the
framework of BD gravity. The motivation for this study comes from
the fact that in string theory, gravity becomes scalar-tensor in
nature which its low energy limit leads to the Einstein gravity,
coupled non-minimally to a scalar field \cite{Green}. Besides, the
ADE and NADE energy densities belong to a dynamical cosmological
constant, thus we need a dynamical frame to accommodate they
instead of Einstein gravity. The investigation on the ADE and NADE
models in the framework of BD cosmology, have been carried out in
\cite{Karami}.

On the other side, there are also several observations which
indicate that the possibility of a mutual interaction between the
DM and DE is not zero. It was argued that the mutual interaction
may solve the coincidence problem \cite{wang2016}. On the other
hand, the simplest form of this mutual interaction can be written
as $Q =3b^2 H(\rho_{m}+\rho_{D})$. Clearly, this interaction is
always positive and hence cannot change itself sign. Considering
the latest observational data, Cai and Su  \cite{Cai2010},
discussed that the sign of the interaction between DM and DE can
change in the redshift around $0.45\leq z\leq 0.9$. Motivated by
\cite{Cai2010}, Wei proposed a sign-changeable interaction term as
$Q=q(\alpha \dot{\rho}+3\beta H{\rho})$, where $\alpha$ and
$\beta$ are dimensionless constant, $q$ is the deceleration
parameter and $\rho$ is the energy density of DE, DM or the sum of
them \cite{WEI2011,Wei2011}. Clearly, the sign of $Q$ is changed
when the expansion of our Universe changes from deceleration $(q >
0)$ to acceleration $(q < 0)$. DE models with sign-changeable
interaction term between two dark sectors have been carried out in
\cite{Abdollahi2016,Abdollahi2017}.

In the present work, we would like to investigate the ADE and NADE
models with sign-changeable interaction term in the background of
BD theory. At first, we study the cosmological implications of
these models and then we perform the stability analysis by
calculating the squared of sound speed $v_s^2={dP}/{d\rho}$
\cite{Peebles20031}. When $v_s^2>0$ we have the classical
stability of a given perturbation. In the framework of Einstein
gravity instability of DE models have been studied in
\cite{StaDE}. While, stability of interacting HDE with GO cutoff
in BD theory has been investigated in \cite{Khodam2014}, sound
instability of nonlinearly interacting ghost dark energy have been
discussed in \cite{Golchin2016}. Recently, we have studied the
stability of the HDE model with the sign-changeable interaction in
BD theory with various IR cutoffs \cite{majid}.

We organize the paper as follows. In section \ref{GF}, we give a
brief review of the interacting ADE model in context of BD
cosmology. In section \ref{ADE} and \ref{NADE}, we investigate
ADE and NADE in the framework of BD theory by assuming a
sign-changeable interaction term, respectively. In each cases, the
cosmological implications of the model as well as the squared
sound stability ${v}^{2}_s$ of the model are studied. Finally, the
summary of the result is discussed in the last section.
\section{Interacting ADE in BD cosmology}\label{GF}
We begin with the action of BD theory, with one scalar field
$\phi$  which in the canonical form can be written \cite{Arik2006}
\begin{equation}
 S=\int{
d^{4}x\sqrt{g}\left(-\frac{1}{8\omega}\phi ^2
{R}+\frac{1}{2}g^{\mu \nu}\partial_{\mu}\phi \partial_{\nu}\phi
+L_M \right)},\label{act1}
\end{equation}
where $\omega$ represents a coupling between scalar field and
gravity, $g$ the determinate of metric tensor $g_{\mu \nu}$, $L_M$
the matter part of the lagrangian, $R$ is the scalar curvature and
$\phi$ is the BD scalar field which replaces with the
Einstein-Hilbert term ${R}/{G}$ in such a way that
$G^{-1}_{\mathrm{eff}}={2\pi \phi^2}/{\omega}$. $G_{\mathrm{eff}}$
is the effective gravitational constant as long as the dynamical
scalar field $\phi$ varies slowly. In order to study the evolution
of the universe, we assume a homogeneous and isotropic FRW
spacetime which is described by the line element
\begin{eqnarray}
ds^2=dt^2-a^2(t)\left(\frac{dr^2}{1-kr^2}+r^2d\Omega^2\right).
\label{metric1}
\end{eqnarray}
where $a(t)$ is the scale factor and $k$ is the curvature
parameter. Since a closed universe with a small positive curvature
($\Omega_k\simeq0.01$) is compatible with observations \cite{spe},
from three possible values $k = -1, 0, 1$, which represent to
open, flat and closed geometry of the universe, we select case $k
=+1$. The variation of the action (\ref{act1}) with respect to the
metric (\ref{metric1}) for universe filled with dust and ADE
yields the following field equations
\begin{eqnarray}
&&\frac{3}{4\omega}\phi^2\left(H^2+\frac{k}{a^2}\right)-\frac{1}{2}\dot{\phi}
^2+\frac{3}{2\omega}H
\dot{\phi}\phi=\rho_m+\rho_D,\label{FE1}\\
&&\frac{-1}{4\omega}\phi^2\left(2\frac{{\ddot{a}}}{a}+H^2+\frac{k}{a^2}\right)-\frac{1}{\omega}H
\dot{\phi}\phi -\frac{1}{2\omega}
\ddot{\phi}\phi-\frac{1}{2}\left(1+\frac{1}{\omega}\right)\dot{\phi}^2=p_{D},\label{FE2}\\
&&\ddot{\phi}+3H
\dot{\phi}-\frac{3}{2\omega}\left(\frac{{\ddot{a}}}{a}+H^2+\frac{k}{a^2}\right)\phi=0,
\label{FE3}
\end{eqnarray}
where the dot is the derivative with respect to time and $\rho_m$
and $\rho_D$ denote the energy density of DM and DE, respectively,
also $H=\dot{a}/a$ is the Hubble parameter and $p_D$ is the
pressure of DE. Furthermore, we exclude baryonic matter and
radiation due to their negligible contribution to the total energy
budget in the late time evolution. Based on the previous
experiences in the BD theory, let us assume the relation between
BD scalar field and scale factor as a power law of the scale
factor, $\phi=\phi_0 a^{\alpha}(t)$. Thus, we have
\begin{equation}
\frac{\dot{\phi }}{\phi }=\alpha H,\   \   \   \   \frac{\ddot{
\phi }}{\phi } =\alpha^{2}H^{2}+\alpha\dot{H},\   \   \   \
\frac{\ddot{\phi }}{\dot{\phi }}=\left(\alpha+\frac{
\dot{H}}{H^{2}}\right)H. \label{square}
\end{equation}
Observational evidences provided by the galaxy cluster Abell
$A586$ supports the interaction between DE and DM
\cite{Bertolami}. In the presence of interaction, the
semi-conservation equations for DE and DM are given by
\begin{eqnarray}
&&
\dot{\rho}_D+3H\rho_D(1+w_D)=-Q,\label{cons1}\\
&&\dot{\rho}_{m}+3H\rho_{m}=Q, \label{cons2}
\end{eqnarray}
where $w_D$ is the equation of state parameter of DE and $Q$ is
the interaction term which we assume has the form $Q =3b^2 q
H(\rho_{m}+\rho_{D})$ \cite{Wei2011,chimen1,chimen2}, $ b^2$ is a
coupling constant and $q$ is the deceleration parameter,
\begin{equation}\label{deceleration}
q=-\frac{\ddot{a}}{a H^2}=-1-\frac{\dot{H}}{H^2}.
\end{equation}
The energy density of ADE in standard cosmology is given by Eq.
(\ref{rhoD}), where the age of universe is defined as
\begin{equation}
T=\int_0^a{dt}=\int_0^a{\frac{da}{Ha}}.
\end{equation}
In the framework of BD cosmology, we write down the energy density
of ADE as
\begin{eqnarray}\label{rhoDBD}
\rho_{D}=\frac{3 n^2\phi^2 }{4\omega T^2},
\end{eqnarray}
where for the NADE we should replace $T$ with $\eta$. The critical
energy density $\rho_{\mathrm{cr}}$ and the energy density of the
curvature $\rho_k$ are introduced as
\begin{eqnarray}\label{rhocr}
\rho_{\mathrm{cr}}=\frac{3\phi^2 H^2}{4\omega},\hspace{0.8cm}
\rho_k=\frac{3k\phi^2}{4\omega a^2}.
\end{eqnarray}
Then the dimensionless density parameters can be written
\begin{eqnarray}
\Omega_m&=&\frac{\rho_m}{\rho_{\mathrm{cr}}}=\frac{4\omega\rho_m}{3\phi^2
H^2}, \label{Omegam} \\
\Omega_k&=&\frac{\rho_k}{\rho_{\mathrm{cr}}}=\frac{k}{H^2 a^2},\label{Omegak} \\
\Omega_D&=&\frac{\rho_D}{\rho_{\mathrm{cr}}}=\frac{4\omega\rho_D}{3\phi^2
H^2}. \label{OmegaD}
\end{eqnarray}
Based on these definitions, and using Eqs.(\ref{square}) and
(\ref{rhocr}), the first Friedmann equation (\ref{FE1}) can be
rewritten as
\begin{equation}\label{rhocr1}
\rho_{\rm cr}+\rho_{k}=\rho_{m}+\rho_{D}+\rho_{\phi},
\end{equation}
where we have defined
\begin{equation}\label{rhocr2}
\rho_{\phi}\equiv\frac{1}{2} \alpha H^2 \phi ^2
\left(\alpha-\frac{3}{\omega}\right).
\end{equation}
Dividing Eq.(\ref{rhocr1}) by $\rho_{\rm cr}$, this equation can be rewritten as
\begin{equation}\label{OmegaD1}
\Omega_{m}+\Omega_{D}+\Omega_{\phi}=1+\Omega_{k},
\end{equation}
where
\begin{equation}\label{Omegaphi}
\Omega_{\phi}=\frac{\rho_{\phi}}{\rho_{\rm cr}}=-2\alpha
\left(1-\frac{\alpha\omega}{3}\right).
\end{equation}
We introduce the ratio of the energy densities as,
\begin{equation}\label{r}
r=\frac{\Omega_{m}}{\Omega_{D}}=-1+\frac{1}{\Omega_{D}}\left[1+\Omega_{k}+2\alpha
\left(1-\frac{\alpha \omega}{3}\right)\right].
\end{equation}
Next, we introduce our approach for investigating the stability of
ADE and NADE in BD theory with sign-changeable interaction against
perturbations. Assuming a small fluctuation in the background of
the energy density, we would like to check whether the
perturbation will grow with time or it propagates as a sound wave
in the medium. In classical perturbation theory, if we consider
$\rho(t)$ as an unperturbed background energy density, then the
perturbed energy density of the back ground in the linear
perturbation factor, can be written as
\begin{equation}\label{pert1}
\rho(t,x)=\rho(t)+\delta\rho(t,x),
\end{equation}
which its energy conservation equation,($\nabla_{\mu}T^{\mu\nu}=0$) yields
\cite{Peebles20031}
\begin{equation}\label{pert2}
\delta\ddot{\rho}=v_s^2\nabla^2\delta\rho(t,x),
\end{equation}
where $v_s^2={dP}/{d\rho}$ is the square of the sound speed. For
case $v_s^2>0$, Eq.(\ref{pert2}) becomes an ordinary wave equation
which have a wave solution in the form $\delta
\rho=\delta\rho_0e^{-i\omega t+i\vec{k}.\vec{x}}$. Obviously it
show a propagation mode for the density perturbations and system
is stable. For case $v_s^2<0$, the frequency of the oscillations
becomes pure imaginary and density perturbations will grow with
time as $\delta \rho=\delta\rho_0e^{\omega t+i\vec{k}.\vec{x}}$
and system cannot be stable. The quantity $v_s^2$ for a nonflat
FRW universe is obtained as
\begin{equation}\label{stable}
{v}^{2}_{s}=\frac{\dot P}{\dot\rho}
=\frac{\dot\rho_{D}w_D+\rho_{D}\dot
{w_D}}{\dot\rho_{D}(1+r)+\rho_{D}\dot r},
\end{equation}
where $P=P_D$ is the pressure of DE and $\rho=\rho_{m}+\rho_{D}$
is the total energy density of DE and DM.
\section{Sign-Changeable ADE in BD theory}\label{ADE}
We begin with the ADE in BD theory, whose energy density is given
by Eq.(\ref{rhoDBD}). Differentiating the expression of the energy
density of ADE given in Eq.(\ref{rhoDBD}) and using
Eqs.(\ref{square}) and (\ref{OmegaD}) we obtain
\begin{eqnarray}
\dot{\rho}_D=2H\rho_D\left(\alpha-\frac{\sqrt{\Omega_D}}{n}\right)\label{rhodota}.
\end{eqnarray}
Inserting this equation in the semi-conservation law (\ref{cons1})
and using Eq.(\ref{r}), we obtain the equation of state parameter
for sign-changeable ADE in BD theory
\begin{eqnarray}\label{wDage}
w_D&=&-1-\frac{2\alpha}{3}+\frac{2}{3n}\sqrt{\Omega_D}-b^2 q (1+r).
\end{eqnarray}
Dividing Eq. (\ref{FE2}) by $H^2$  and using Eqs. (\ref{square}),
(\ref{deceleration}), (\ref{rhoDBD}), (\ref{Omegak}) and
(\ref{OmegaD}) we can obtain the following expression for the
deceleration parameter
\begin{equation}\label{deceleration1}
q=\frac{1}{2\alpha+2}\left[(2\alpha+1)^2+2\alpha (\alpha \omega
-1)+3 \Omega_{D} w_{D}\right].
\end{equation}
Substituting $w_D$ from Eq. (\ref{wDage}) in the above relation,
we arrive at
\begin{equation}\label{deceleration2}
q=
\frac{1+2\alpha(1+\alpha(2+\omega))+\Omega_k-(3+2\alpha)\Omega_D+
\frac{2{\Omega_D}^{{3}/{2}}}{n}}{2(1+\alpha)+3b^2 \Omega_D (1+r)}.
\end{equation}
The equation of motion for ADE may be obtained by substituting Eq.
(\ref{rhoDBD}) in Eq.(\ref{OmegaD}). We find
\begin{eqnarray}\label{OmegaD1}
\Omega_D&=&\frac{n^2}{H^2 T^2}.
\end{eqnarray}
Taking the time derivative of Eq. (\ref{OmegaD1}) and using Eq.
(\ref{deceleration}) as well as the fact that
$\dot\Omega_{D}={\Omega}^{\prime}_{D} H$,  we get
\begin{eqnarray}\label{OmegaDa}
{\Omega'_D}=2\Omega_D\left(1+q-\frac{\sqrt{\Omega_D}}{n}\right),
\end{eqnarray}
where the prime denotes derivative with respect to $x=\ln a$.
Stability of this model can be studied by taking derivative of
Eq.(\ref{wDage}) and using Eqs.(\ref{rhodota}) and (\ref{stable}).
Since the expression of ${v}^{2}_{s}$ is too long, for the
economic reason we do not present it here, instead we focus on its
behaviour via figures.
\begin{figure}[htp]
\begin{center}
\includegraphics[width=8cm]{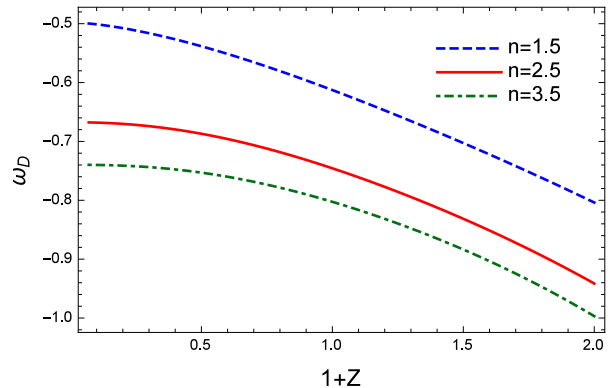}
\caption{Evolution of $w_D$ versus redshift parameter $z$ for the
sign-changeable interacting ADE in BD cosmology. Here, we have
taken $\alpha=0.003$, $\omega=10^4$ ,$\Omega_k=0.01$ and $b^2=0.1$
as the initial condition. }\label{EoS-z1}
\end{center}
\end{figure}

\begin{figure}[htp]
\begin{center}
\includegraphics[width=8cm]{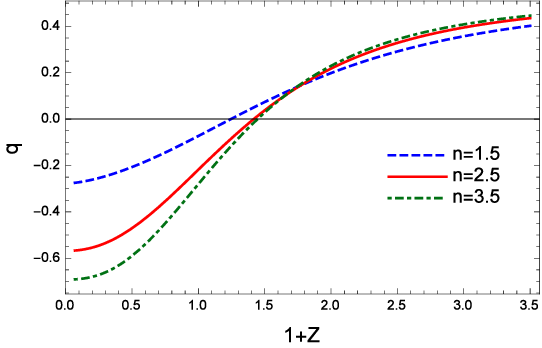}
\caption{Evolution of the deceleration parameter $q$ against
redshift parameter $z$ for the sign-changeable interacting ADE in
BD cosmology. Here, we have taken $\alpha=0.003$, $\omega=10^4$
,$\Omega_k=0.01$ and $b^2=0.1$ as the initial condition.
}\label{q-z1}
\end{center}
\end{figure}

\begin{figure}[htp]
\begin{center}
\includegraphics[width=8cm]{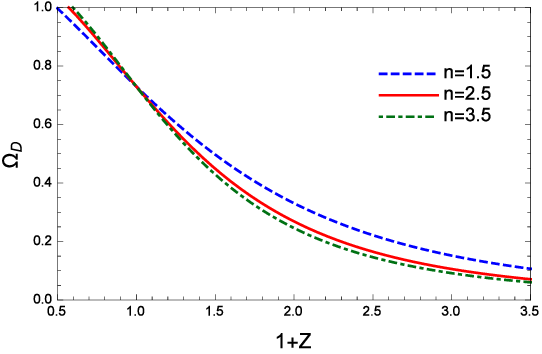}
\caption{Evolution of $\Omega_D$ versus redshift parameter $z$ for
the sign-changeable interacting ADE in BD cosmology. Here, we have
taken $\alpha=0.003$, $\omega=10^4$ ,$\Omega_k=0.01$ and $b^2=0.1$
as the initial condition. }\label{Omega-z1}
\end{center}
\end{figure}

\begin{figure}[htp]
\begin{center}
\includegraphics[width=8cm]{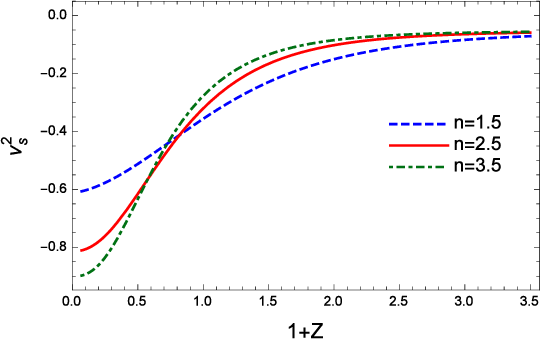}
\includegraphics[width=8cm]{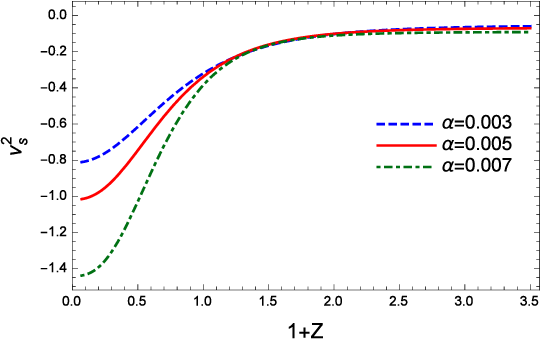}
\caption{Evolution of the squared of sound speed ${v}^{2}_{s}$
against redshift parameter $z$ for the sign-changeable interacting
ADE in BD cosmology. Here, we have taken $\alpha=0.003$,
$\omega=10^4$ ,$\Omega_k=0.01$ and $b^2=0.1$ in the left panel and
$\Omega_k=0.01$, $\omega=10^4$, $n=2.5$ and $b^2=0.1$  in the
right panel, as the initial condition, respectively}\label{S-z1}
\end{center}
\end{figure}

\begin{figure}[htp]
\begin{center}
\includegraphics[width=8cm]{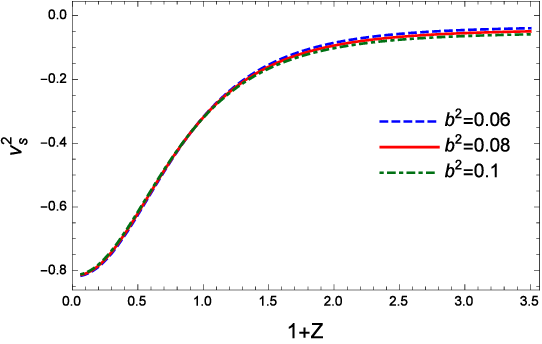}
\includegraphics[width=8cm]{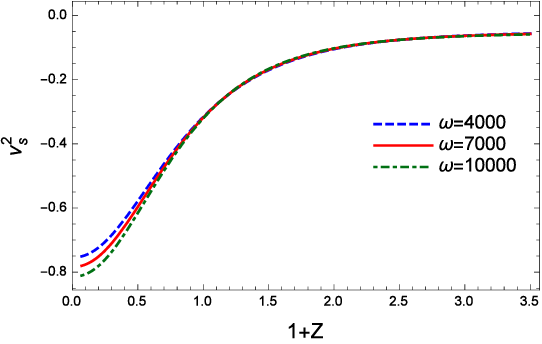}
\caption{Evolution of the squared of sound speed ${v}^{2}_{s}$
against redshift parameter $z$  for the sign-changeable
interacting ADE in BD cosmology. Here, we have taken
$\alpha=0.003$, $\omega=10^4$ ,$\Omega_k=0.01$ and $n=2.5$ in the
left panel and  $\alpha=.003$, $n=2.5$  ,$\Omega_k=0.01$ and
$b^2=0.1$  in the right panel, as the initial condition,
respectively}\label{S-z2}
\end{center}
\end{figure}
To describe the evolution of the universe, we plot the
cosmological parameters for the sign-changeable interacting ADE in
BD cosmology. From Fig.~\ref{EoS-z1}, we see that $w_D$ cannot
cross the phantom line, while according to Fig.~\ref{q-z1}, we see
that the deceleration parameter $q$ transits from deceleration
($q>0$) in the early time to acceleration ($q< 0$) in the last
time around $z\approx 0.6$. Again, by keeping the same initial
condition, we plot the evolution of $\Omega_D$ against redshift
parameter in Fig.~\ref{Omega-z1} which show that at the late time
where the DE is dominated we have $\Omega_D\rightarrow 1$, while
$\Omega_D\rightarrow 0$ at the early time. Finally, we plot the
squared sound speed for ADE model in BD theory in Figs.~\ref{S-z1}
and \ref{S-z2} by considering the different parameters $\alpha$,
$b^2$, $n$ and $\omega$. In Fig.~\ref{S-z1}, we plot ${v}^{2}_{s}$
versus $z$ with different values of $n$ also $\alpha$, which show
we cannot have the stable model. According to Fig.~\ref{S-z2}, we
see that for different values of  $b^2$ and $\omega$, the model
does not show a signal of stability.
\section{Sign-Changeable NADE in BD theory}\label{NADE}
Since the original model of ADE model suffers the difficulty to
describe the matter-dominated epoch, the NADE was proposed by Wei
and Cai \cite{Cai20072} to describe the late time acceleration. In
the NADE model, the conformal time $\eta$ is chosen as the cutoff
instead the age of the universe, which leads to the energy density
in the form \cite{Cai20072}
\begin{equation}\label{rhonage}
\rho_{D}= \frac{3n^2 m_{p}^2}{\eta^2},
\end{equation}
where the conformal time is given by
\begin{equation}
\eta=\int_0^a{\frac{da}{Ha^2}}.
\end{equation}
In the framework of BD cosmology, by using Eqs. (\ref{rhoD}) and
(\ref{OmegaD}) we can write the energy density of NADE as
\begin{equation}\label{rho1n}
\rho_{D}= \frac{3n^2\phi^2 }{4\omega \eta^2},
\end{equation}
and
\begin{equation}\label{OmegaDa1}
\Omega_D=\frac{n^2}{H^2\eta^2}.
\end{equation}
Differentiating Eq.(\ref{rho1n}), we arrive at
\begin{eqnarray}
\dot{\rho}_D=2H\rho_D\left(\alpha-\frac{\sqrt{\Omega_D}}{na}\right)\label{rhodotna}.
\end{eqnarray}
Substituting this relation in Eq. (\ref{cons1}), after using
Eq.(\ref{r}), we find
\begin{eqnarray}\label{wDnage}
w_D&=&-1-\frac{2\alpha}{3}+\frac{2}{3na}\sqrt{\Omega_D}-b^2 q (1+r).
\end{eqnarray}
When $q=1$, Eq.(\ref{wDnage}) restores the equation of state of
the NADE in the BD theory  \cite{Skheykhiage}. Setting $\alpha=0$
and $q=1$, this equation recovers its respective expression for
interacting NADE in Einstein gravity \cite{shey0}. We can also
obtain the deceleration parameter $q$ by substituting
Eq.(\ref{wDnage}) in Eq.(\ref{deceleration1}). The result is
\begin{equation}\label{deceleration3}
q= \frac{1+2\alpha(1+\alpha(2+\omega))+\Omega_k-
(3+2\alpha)\Omega_D+\frac{2{\Omega_D}^{\frac{3}{2}}}{na}}{2(1+\alpha)+3b^2
\Omega_D (1+r)}.
\end{equation}
On the other hand, the equation of motion for $\Omega_D$ takes the
form
\begin{eqnarray}\label{OmegaDna}
{\Omega'_D}=2\Omega_D\left(1+q-\frac{\sqrt{\Omega_D}}{na}\right).
\end{eqnarray}
Finally, we investigate stability of NADE by calculating
${v}^{2}_{s}$. For the economic reason, we do not bring the
explicit expression for ${v}^{2}_{s}$, instead we study the
evolution of ${v}^{2}_{s}$ via figures.
\begin{figure}[htp]
\begin{center}
\includegraphics[width=8cm]{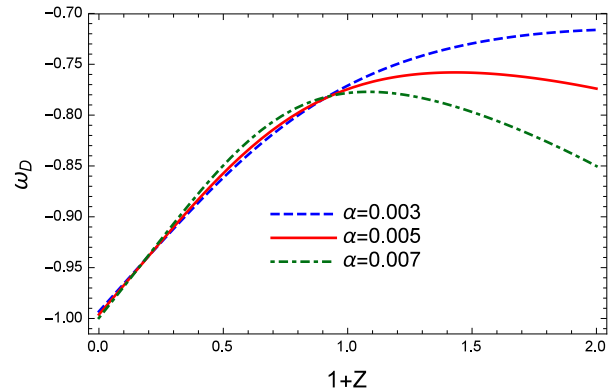}
\caption{Evolution of $w_D$ versus redshift parameter $z$ for the
sign-changeable interacting NADE in BD cosmology. Here, we have
taken $n=2.5$, $\omega=10^4$ ,$\Omega_k=0.01$ and $b^2=0.01$ as
the initial condition. }\label{EoS-z2}
\end{center}
\end{figure}

\begin{figure}[htp]
\begin{center}
\includegraphics[width=8cm]{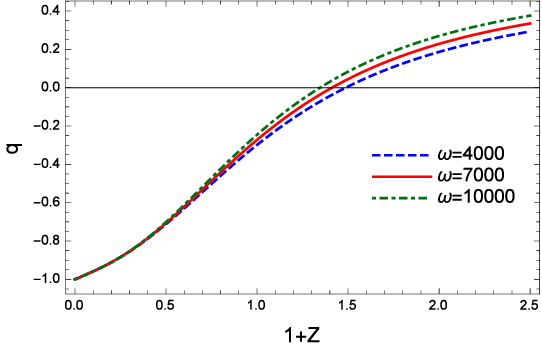}
\caption{Evolution of the deceleration parameter $q$ against
redshift parameter $z$ for the sign-changeable interacting NADE in
BD cosmology. Here, we have taken $n=2.5$, $\alpha=0.003$
,$\Omega_k=0.01$ and $b^2=0.01$ as the initial condition.
}\label{q-z2}
\end{center}
\end{figure}

\begin{figure}[htp]
\begin{center}
\includegraphics[width=8cm]{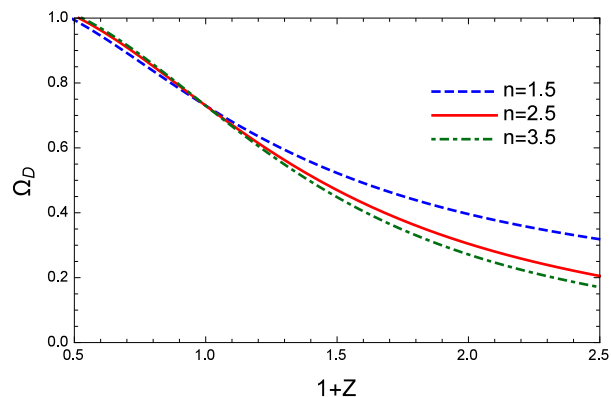}
\caption{Evolution of $\Omega_D$ versus redshift parameter $z$ for
the sign-changeable  interacting NADE in BD cosmology. Here, we
have taken $\omega=10^4$, $\alpha=0.003$ ,$\Omega_k=0.01$ and
$b^2=0.01$. }\label{Omega-z2}
\end{center}
\end{figure}

\begin{figure}[htp]
\begin{center}
\includegraphics[width=8cm]{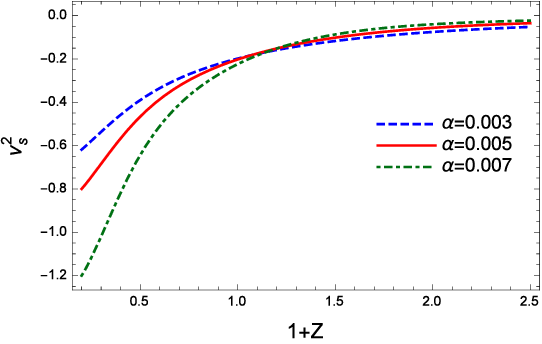}
\includegraphics[width=8cm]{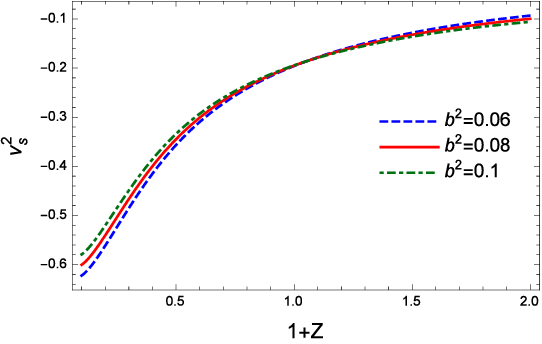}
\caption{Evolution of the squared of sound speed ${v}^{2}_{s}$
against redshift parameter $z$ for the sign-changeable interacting
NADE in BD cosmology. Here, we have taken $n=2.5$, $\omega=10^4$
,$\Omega_k=0.01$ and $b^2=0.01$ in the left panel and
$\alpha=0.003$, $\omega=10^4$, $\Omega_k=0.01$ and $n=2.5$  in the
right panel, as the initial condition, respectively}\label{S-z11}
\end{center}
\end{figure}

\begin{figure}[htp]
\begin{center}
\includegraphics[width=8cm]{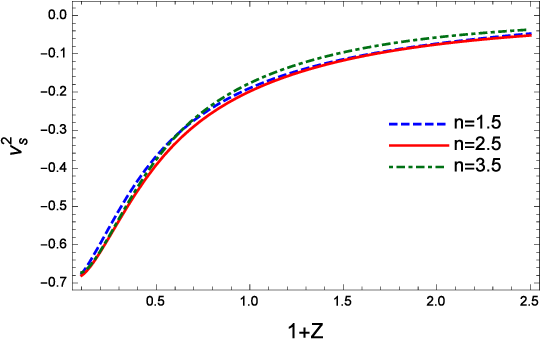}
\includegraphics[width=8cm]{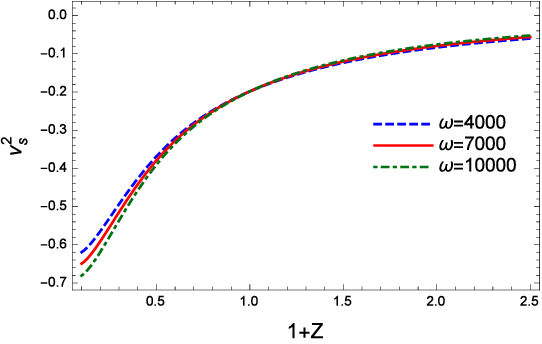}
\caption{Evolution of the squared of sound speed ${v}^{2}_{s}$
against redshift parameter $z$  for the sign-changeable
interacting NADE in BD cosmology for $\alpha=0.003$, ,
$\Omega_k=0.01$ and $b^2=0.01$.  As the initial condition, we have
taken in the left panel $\omega=10^4$, and  in the right panel
$n=2.5$.}\label{S-z22}
\end{center}
\end{figure}
The behaviors of $w_D$, $q$ and $\Omega_D$ against redshift
parameter $z$ are plotted in Figs.~\ref{EoS-z2}-\ref{Omega-z2}.
Our analysis of these figures show that $w_D$ cannot cross phantom
line. From Fig.~\ref{q-z2}, we see that a deceleration phase ends
at past and transits to a phase of acceleration at late time. In
Fig.~\ref{Omega-z2}, which we plot $\Omega_D$ versus $z$ for the
sign-changeable interacting NADE in BD cosmology, we see
$\Omega_D\rightarrow 1$ at late time for different values of $n$.
Finally, in Figs.~\ref{S-z11} and \ref{S-z22}, the stability of
model is studied which show for different values of parameters
which confirm that this system cannot show signal of stability in
our universe.
\section{Closing remarks}
In this paper, we have considered the ADE and NADE with
sign-changeable interaction term in the framework of BD theory. We
have discussed the physical behavior of the EoS parameter, the
deceleration parameter, the evolution of density parameter
$\Omega_D$ and the squared sound speed ${v}^{2}_{s}$ versus the
redshift parameter $z$ for ADE and NADE in BD cosmology with
sign-changeable interaction term. For ADE model, we found out that
the EoS parameter $w_D$ cannot cross the phantom line, while the
deceleration parameter $q$ transits from deceleration ($q>0$) in
the early time to acceleration ($q< 0$) at the last time around
$z\approx 0.6$ which is compatible with recent observations. We
also plotted the evolution of ${v}^{2}_{s}$ versus $z$ in and
observed that this model cannot lead to stable DE dominated
universe. For NADE model, we again see that $w_D$ cannot cross the
phantom line, while at late time where the DE dominates we have
$\Omega_D\rightarrow 1$. Finally, we observed that ${v}^{2}_{s}$
remains negative so we have a sign of instability for the NADE in
BD theory. In conclusion, our studies show that for the
sign-changeable ADE and NADE models in the set up of BD cosmology
we cannot have a stable DE dominated universe.
\acknowledgments{We thank Shiraz University Research Council. This
work has been supported financially by Research Institute for
Astronomy \& Astrophysics of Maragha (RIAAM), Iran.}


\end{document}